\documentclass{llncs}
\usepackage{graphicx}
\usepackage{float}
\newfloat{example}{h}{ttl}[section]
\floatname{example}{Example}

\begin{document}

\title{Provenance and evidence in UniProtKB}
\author{Jerven Bolleman \and Alain Gateau \and Sebastien Gehant \and Nicole Redaschi \and the UniProt Consortium}
\institute{Swiss Institute of Bioinformatics
\linebreak
\email{jerven.bolleman@isb-sib.ch},
\email{alain.gateau@isb-sib.ch},
\email{sebastien.gehant@isb-sib.ch},
\email{nicole.redaschi@isb-sib.ch},}
\maketitle

\begin{abstract}

The primary mission of UniProt is to support biological research by maintaining a stable, comprehensive, fully classified, richly and accurately annotated protein sequence knowledgebase, with extensive cross-references to external resources, that is freely available to the scientific community. 

To enable users of the knowledgebase to accurately assess the reliability of the information contained in this resource, the evidence for and provenance of the information must be recorded. This paper discusses the user requirements for this kind of metadata and the manner in which UniProt\-KB records it.

\end{abstract}

\section{The UniProt Knowledgebase}
\label{sec:uniprotkb}

The UniProt Knowledgebase (UniProtKB)~\cite{uniprot} consists of two parts: \discretionary{UniProtKB/}{Swiss-Prot}{UniProtKB/Swiss-Prot}, containing manually annotated records describing proteins with information extracted from the scientific literature and curator-evaluated computational analysis, and \discretionary{UniProtKB/}{TrEMBL}{UniProtKB/TrEMBL}, with automatically annotated records. The UniProtKB is available in several data formats, including RDF \cite{rdf}, at \url{www.uniprot.org}. The RDF distribution currently consists of almost 2.3 billion triples\footnotemark, making it one of the largest data sets that is available in RDF. Roughly 10\% of these triples are metadata that provide provenance and evidence information and this percentage is expected to increase.

\footnotetext{Data for UniProt release 2010\_11 of 2-11-2010}

\section{A brief history of provenance}

In general, scientists tend to have higher confidence in information that has been reviewed and approved by other scientists than in unreviewed data that was predicted by programs. In UniProtKB this is reflected by a split into two main sections: the manually curated and reviewed \discretionary{UniProtKB/}{Swiss-Prot}{UniProtKB/Swiss-Prot} and the automatically annotated and unreviewed \discretionary{UniProtKB/}{TrEMBL}{UniProtKB/TrEMBL}.

But even manually reviewed data comes with different degrees of reliability. Historically, only four basic types of evidences were distinguished:

\begin{description}
\item [Experimental] The annotation is supported by experimental evidence.
\item [Probable] The annotation was inferred with good confidence by a curator from another annotation with experimental evidence (e.g. an experimentally proven transcription factor is inferred to be located in the cell nucleus).
\item [By similarity] The annotation was inferred based on a high sequence similarity to a protein which has experimental evidence for the same annotation.
\item [Potential] The annotation was predicted by a sequence analysis tool.
\end{description}

This course-grained approach no longer meets the demands of bioinformaticians who ask for more detailed evidence information and, over all, provenance information. UniProt hence extended the metadata it records to include the following:

\begin{description}
\item[Source] Indicates the source of the information. There are three main categories:

\begin{description}
\item[Scientific literature] UniProtKB/Swiss-Prot records summarize the published knowledge about a protein in a condensed way. Scientists interested in more detailed information can retrieve and read the original publications. 

\item[Programs] Some UniProtKB annotations are inferred from the results of various sequence analysis programs. Distinguishing these annotations from those with an experimental origin is important for users using UniProtKB as a training set for prediction programs.

\item[External databases] UniProtKB also imports data from other databases. Scientists interested in more detailed information can retrieve and read the external database records. Can also be used internally to synchronize information.
\end{description}

\item[Date] Indicates when a statement has last been updated. Can also be used internally to select aging records for which annotation might need to be revised.
\end{description}

UniProt has also extended its evidence codes slightly, and is currently investigating the adoption of the GO evidence codes~\cite{go}, which are widely used in the biocuration field, to provide even more detailed and standardized evidence information.

\section{Provenance representation}

We will use the semi-automatic annotation program HAMAP \cite{hamap} to illustrate how provenance information is currently described in the UniProtKB. This program uses manually curated protein family rules to annotate all proteins that belong to the same family. In the examples shown, it has annotated the recommended protein name based on the family rule MF\_00536. In the RDF representation, we use reification to attach the provenance information to the name of the protein (see example \ref{rdf:provenance} in turtle syntax with prefixes omitted for readability). Using SPARQL, one can retrieve all proteins whose names were annotated by this HAMAP rule (see example \ref{sparql:current}).

\begin{example}{}
\begin{verbatim}
protein:Q65EJ5 :recommendedName _:0 .
_:0 rdf:type :Structured_Name ;
    :fullName "4-hydroxythreonine-4-phosphate dehydrogenase" .
_:1 rdf:type rdf:Statement ;
    rdf:Subject _:0 ;
    rdf:Predicate :fullName ;
    rdf:Object "4-hydroxythreonine-4-phosphate dehydrogenase" ;
    :attribution _:2 .
protein:Q65EJ5 :attribution _:2 .
    :source hamap:MF_00536 ;
    :date "2010-08-04"
\end{verbatim}
\caption{RDF representation of the source of a protein name.}
\label{rdf:provenance}
\end{example}

\begin{example}{}
\begin{verbatim}
select 
  ?protein ?date ?predicate ?object
where {
?reif rdf:subject ?subject ;
  rdf:predicate ?predicate ;
  rdf:object ?object ;
  :attribution ?attribution .
?protein :attribution ?attribution .
?attribution :source hamap:MF_00536 ;
  :date ?date     . }
\end{verbatim}
\caption{Finding annotations that where added by HAMAP rule MF\_00536 using SPARQL.}
\label{sparql:current}
\end{example}

UniProtKB data is also distributed in plain XML. Since the XML specification does not define a standard mechanism for attaching provenance information, we had to invent our own solution (see example \ref{xml:provenance}). Evidence and provenance information is stored in "evidence" elements. The data to which the evidence applies links to its evidence via an "evidence" attribute. This means that users who are interested in this information have to write custom parsers to link the "evidence" attribute values to the corresponding "key" attributes of the "evidence" elements.

\begin{example}{}
\begin{verbatim}
<recommendedName ref="1">
<fullName evidence="EA5">4-hydroxythreonine-4-phosphate dehydrogenase</fullName>
</recommendedName>
<evidence key="EA5" category="import" type="HAMAP" attribute="MF_00536" date="2010-08-04" />
\end{verbatim}
\caption{XML representation of the source of a protein name.}
\label{xml:provenance}
\end{example}

\section{Conclusions \& further work}

Users in the bioinformatics community demand per statement provenance. Only the RDF data format can deliver this in a standardized and effective manner. RDF reification provides the low level infrastructure needed for adding provenance to a statement. Unfortunately, the reification syntax is awkward to use. A simple improvement such as having a shorthand in SPARQL for the reification quad would help (see example \ref{sparql:suggested}).
 
\begin{example}{}
\begin{verbatim}
select 
  ?protein ?date ?predicate ?object
where {
    reification(?subject :predicate ?object ) ;
    :attribution ?attribution .
?protein :attribution ?attribution ;
?attribution :source hamap:MF_00536 ;
    :date ?date . }
\end{verbatim}
\caption{Suggested function syntax easing the use of the reification quad in SPARQL queries.}
\label{sparql:suggested}
\end{example}

\section{Acknowledgments}

This activity at SIB Swiss Institute of Bioinformatics is mainly supported by the Swiss Federal Government through the Federal Office of Education and Science, by the National Institutes of Health (NIH) grant 2 U01 HG02712-04,  and the European Commission contract SLING (226073).

\end{document}